\documentclass[prl,showkeys,twocolumn,longbibliography]{revtex4-1}
\usepackage[utf8]{inputenc}
\usepackage{graphicx}
\usepackage{textcomp}
\usepackage{txfonts}
\usepackage{url}
\urldef{\SMonline}\url{https://journals.aps.org/prb}
\usepackage{color}
\usepackage{gensymb} % ° degree symbol
\usepackage{tikz}
\usetikzlibrary{positioning} % arrows in math
\usetikzlibrary{calc} % arrows in math
\usetikzlibrary{shapes}
\usetikzlibrary{arrows}

\newcommand{\dublin}{School of Physics, AMBER and CRANN Institute, Trinity College, Dublin 2, Ireland}

\newcommand{\eqref}[1]{(\ref{#1})}

\usepackage{xcite}
\externalcitedocument{main}

\begin{document}

\title{Erratum: Photovoltage from ferroelectric domain walls in BiFeO$_3$}

\author{Sabine K\"{o}rbel}
%\affiliation{\dublin,\prague}
\affiliation{\dublin}
\email{skoerbel@uni-muenster.de}
\author{Stefano Sanvito}
\affiliation{\dublin}

\setcounter{figure}{-1} 

\begin{abstract}
In the original article a mistake in the methodology lead to an incorrect prediction of exciton delocalization below
a critical exciton density. This unrealistic delocalized exciton state yielded a sizable domain-wall photovoltage.
When done correctly, the simulations yield self-trapped (localized) excitons at all considered exciton densities,
without any evidence for a transition to a delocalized exciton. This realistic self-trapped exciton state yields
only a negligible domain-wall photovoltage. The original conclusion that ferroelectric domain walls could
be responsible for the measured photovoltage if carrier lifetime and diffusion length are higher than expected is incorrect.
Correct is: The domain-wall photovoltage in BiFeO$_3$ is much too small to explain the measured
photovoltage. The original analysis is meaningful for ferroelectrics without carrier self-trapping, not for BiFeO$_3$.
%\\ \ \\
%\noindent The revised abstract reads: \\ \ \\
%\noindent
%We calculate the component of the photovoltage in bismuth ferrite that is generated by ferroelectric do-
%main walls, using first-principles methods, in order to compare its magnitude to the experimentally measured
%photovoltage of the order of 10~mV per domain. We find that in principle, a ferroelectric domain wall can
%create a photovoltage of the order of 10~mV, if carrier lifetimes and diffusion lengths are large enough ($\approx$100~$\mu$s 
%and $\approx$100~nm, respectively). In the case of BiFeO$_3$, however, the domain-wall photovoltage is by orders of
%magnitude smaller due to self-trapping of charge carriers.
\end{abstract}
%
%\pacs{}
\maketitle
%
%In this Erratum the structure of the original article is
%maintained, except for additional sectioning. %for the sake of
%clarity. First we describe the error in the original article, how
%it affects the different parts of the original article, and how
%to correct it. More details are then given in the individual
%sections and subsections.
%\\ \ \\ 
\noindent
The mistake in the original article consists in modeling
$M$-fold diluted exciton polarons by means of a fraction $1/M$
of an exciton polaron ($M\gg1$) instead of using an $M$-times
as large simulation box. This fractional-exciton approach
has no predictive power for polarons. It will always predict
vanishing atomic displacements and delocalized excitonic
states for low exciton concentrations. This is because the
atomic displacements that cause carrier self-trapping should be
approximately proportional to the amount of excess charge
$|\psi|_{\mathrm{excess}}^2$ %\cite[51]{sio:2019:polarons},
[see Sio et al., Phys. Rev. Lett. 122, 246403 (2019), Eq.~(3)],
here $|\psi|_{\mathrm{excess}}^2/M$ with $M\gg 1$. When we perform
the simulation correctly with an integer number of charge
carriers in a larger supercell, we observe self-trapping for
all supercell sizes considered (up to 320 atoms, see section
``Self-trapping at low carrier concentrations'' in the Supplemental Material \footnote{See Supplemental Material at \SMonline\ for details \label{footnote:SM}}.
\\ \ \\
\noindent
The original article is affected by the mistake as follows:
\\ \ \\\noindent
\textbf{Incorrect in the original:} The original statement that there
was a transition between self-trapped and delocalized exciton
is incorrect. Figures 3(a), Fig. 4, and Fig. 5 of the original
article are wrong. In the original article, based on a fractional-exciton approach, we wrongly stated
that at sufficiently low exciton densities, the excitons were
delocalized. 
%In this case we neglected changes in atomic
%coordinates caused by the excitons and instead included
%ionic screening by means of an effective dielectric constant.
A fractional-exciton approach is only meaningful for a
hypothetical ferroelectric that does not exhibit charge-carrier
self-trapping. Such a ferroelectric may or may not exist.
BiFeO$_3$ is not such a ferroelectric. %If we want to make predictions for 
For BiFeO$_3$ it is necessary to consider self-trapping of
charge carriers. %While in the original article it was concluded
The original conclusion
that ferroelectric domain walls in BiFeO$_3$ can account for
photovoltages of the order of the experimentally measured
%ones of about
$\approx$10 mV per domain, if carrier lifetimes and
diffusion lengths are large, is incorrect. 
%this is no longer true when carrier self-trapping is considered.
\\ \ \\\noindent
\textbf{Correction:} Part of the original
results obtained without carrier self-trapping are still meaningful.
We assume that before self-trapping of an exciton is
completed, the exciton is delocalized, and the fractional-exciton approach can be used to calculate the domain-wall
photovoltage as it was done in the original article. Once
self-trapping of the exciton is completed, the exciton no
longer contributes to the domain-wall photovoltage because
it has lost its dipole moment,
 see Fig.~\ref{fig:densities_small_large_X}, 
and the
domain-wall photovoltage is approximately zero until the
end of the exciton lifetime. The measurable domain-wall
photovoltage is a time average. When carrier self-trapping
is correctly accounted for, the domain-wall photovoltage of
BiFeO$_3$ is lower than the experimentally measured one by
orders of magnitude, indicating that the measured photovoltage of BiFeO$_3$ should arise from other causes than pristine
ferroelectric domain walls.
\\ \ \\
\noindent
\textbf{Addition:} To determine the time needed for carrier self-trapping, we performed molecular dynamics simulations of
excitons at ferroelectric domain walls.
\\ \ \\
\noindent
The individual sections of the original article are affectd by the mistake as follows:
%%%%%%%%%%%%%%%%%%%%%
\section{Introduction} 
%%%%%%%%%%%%%%%%%%%%%
\noindent\textbf{Correct in the original:} The original Introduction is correct.
\\ \ \\\noindent
\textbf{Addition:} First we consider domain-wall electrostatics
in the ground state, without photo-generated charge carriers,
to determine the potential energy landscape that photocarriers
experience. Next, we add electron-hole pairs, forbidding
changes in the atomic positions (forbidding self-trapping of ex-
citons), and show that in this case the domain walls separate
electron and hole and can give rise to a photovoltage about
as large as the one measured in experiment. This idealized
case corresponds to a BiFeO$_3$-like ferroelectric without
exciton self-trapping, which may or may not exist. Finally,
we consider real BiFeO$_3$ and take into account unavoidable
changes in the atomic coordinates (we allow for spontaneous
self-trapping of the excitons), and show that now the domain
walls no longer separate electron and hole: electron-hole
attraction is stronger. In the self-trapped exciton state (small
exciton polaron), both electron and hole reside on the same
side of the domain wall.

%
%%%%%%%%%%%%%%%%%%%%%%%%%%%%%%%%%%%%%%%%
\section{\label{sec:methods}Methods}
%%%%%%%%%%%%%%%%%%%%%%%%%%%%%%%%%%%%%%%%
\noindent\textbf{Correct in the original:} The original Methods section is
correct.

%%%%%%%%%%%%%%%%%%%%%%%%%%%%%%%%%%%%%%%%
\subsection{Domain walls without excitons}
%%%%%%%%%%%%%%%%%%%%%%%%%%%%%%%%%%%%%%%%
The part of the original Methods section about domain walls without excitons is correct.
%
%%%%%%%%%%%%%%%%%%%%%%%%%%%%%%%%%%%%%%%%
\subsection{Domain walls with excitons}
%%%%%%%%%%%%%%%%%%%%%%%%%%%%%%%%%%%%%%%%

\noindent\textbf{Addition:} We perform two types of calculations with
electron-hole pairs (excitons) at ferroelectric domain walls: \\
\indent 1) Atomic positions fixed. This case corresponds to a hypothetical ferroelectric material which does not exhibit 
self-trapping of carriers. The domain-wall photovoltages obtained
for this hypothetical ferroelectric tell us how large domain-wall photovoltages in ferroelectrics can in principle become.
We calculate the domain-wall photovoltage from the electronic potential and correct it for ionic screening by means
of an effective dielectric constant. Here we consider different exciton concentrations by means of fractional occupations
of electron and hole state. This fractional-exciton approach
is valid only if atomic displacements caused by the exciton
(polaronic effects) can be neglected.\\
\indent 2) Atomic positions optimized (we allow for carrier
self-trapping). The calculations take longer, therefore smaller
supercells are used. These calculations correspond to real
BiFeO$_3$, as far as DFT$+U$ calculations can represent a real
material. These calculations tell us how large the domain-wall
photovoltage can become in BiFeO$_3$ and similar materials
that exhibit carrier self-trapping. Since polaronic effects are
important, here it is necessary to consider an integer number
of excitons in the simulation box.
\\ \ \\\noindent
\textbf{Modification:}
%%%%%%%%%%%%%%%%%%%%%%%%%%%%%%%%%%%%%%%%%%%%%%%%%%%%%%%%%%%%%%%%%
\subsubsection*{Atomic coordinates fixed}
%%%%%%%%%%%%%%%%%%%%%%%%%%%%%%%%%%%%%%%%%%%%%%%%%%%%%%%%%%%%%%%%%

Calculations were performed using supercells with 280
atoms. Exciton concentrations from 0.005 to 0.1 excitons per
280-atom supercell were considered. Only the electronic den-
sity was optimized, cell parameters and atomic positions were
kept fixed to their ground-state values. Hence, self-trapping
of excitons (small exciton polaron formation) was forbidden.

%%%%%%%%%%%%%%%%%%%%%%%%%%%%%%%%%%%%%%%%%%%%%%%%%%%%%%%%%%%%%%%%%
\subsubsection*{Atomic coordinates optimized}
%%%%%%%%%%%%%%%%%%%%%%%%%%%%%%%%%%%%%%%%%%%%%%%%%%%%%%%%%%%%%%%%%

Calculations were performed in supercells containing 120
atoms. We consider a single exciton concentration of one
exciton per 120-atom supercell. Both the atomic positions
and cell parameters were optimized until the energy difference between subsequent ionic relaxation steps fell below
0.1 meV. Hence, self-trapping of excitons (small exciton
polaron formation) was allowed (not enforced). The domain-wall photovoltage was not calculated explicitly because the
electronic potential in the optimized geometry was too noisy
to extract a potential profile. However, visual inspection of the
electron and hole wavefunction (Fig.~\ref{fig:densities_small_large_X}) indicates that the
exciton's dipole moment is negligible compared to the case
with fixed atomic coordinates.
\\ \ \\\noindent
\textbf{Addition:}

%%%%%%%%%%%%%%%%%%%%%%%%%%%%%%%%%%%%%%%%%%%%%%%%%%%%%%%%%%%%%%%%%
\subsubsection{Ab initio molecular dynamics}
%%%%%%%%%%%%%%%%%%%%%%%%%%%%%%%%%%%%%%%%%%%%%%%%%%%%%%%%%%%%%%%%%
To determine the time needed for exciton self-trapping, we
performed first-principles molecular dynamics simulations at
0 K and at 300 K to exclude and include thermal effects. In
the case of room temperature, we first equilibrated the system without excitons for 500~fs. 
The simulations were carried out in the $NVT$ ensemble using an Anderson thermostat
with $\nu$=0.5 (0~K) and $\nu$=0.1 (300~K), respectively, where $\nu$ is the probability for velocity reassignment per atom and time step. 
Supercells containing 120 atoms were employed. The
timestep was set to 1~fs. The supercells are too small and $\nu$
 is too large to approach the thermodynamic limit, therefore the simulated dynamics can only yield
an estimate for the order of magnitude of the time needed for
exciton self-trapping.

%%%%%%%%%%%%%%%%%%%%%%%%%%%%%%%%%%%%%%%%%%
\section{Results and discussion}
%%%%%%%%%%%%%%%%%%%%%%%%%%%%%%%%%%%%%%%%%%
%%%%%%%%%%%%%%%%%%%%%%%%%%%%%%%%%%%%%%%%%%%%%%%%
\subsection{Domain walls without excitons}
%%%%%%%%%%%%%%%%%%%%%%%%%%%%%%%%%%%%%%%%%%%%%%%%
\noindent\textbf{Correct in the original:} The original results for domain walls
without excitons are correct. The first part of the original
Results section and Figures 1 and 2, which show the ferroelectric polarization and electronic potential profile at 71\textdegree~ and
109\textdegree~domain walls, are correct.

%
%%%%%%%%%%%%%%%%%%%%%%%%%%%%%%%%%%%%%%%%%%%%%%%%%%%%%%%%%%%%%%%%%%%%%%%%%%%%%%%%%%%%%%%%%%%%%%%%%%%%%%%%
\subsection{Domain walls with excitons}
%%%%%%%%%%%%%%%%%%%%%%%%%%%%%%%%%%%%%%%%%%%%%%%%%%%%%%%%%%%%%%%%%%%%%%%%%%%%%%%%%%%%%%%%%%%%%%%%%%%%%%%%
\subsubsection{Atomic coordinates fixed}
%%%%%%%%%%%%%%%%%%%%%%%%%%%%%%%%%%%%%%%%%%%%%%%%%%%%%%%%%%%%%%%%%%%%%%%%%%%%%%%%%%%%%%%%%%%%%%%%%%%%%%%%
\noindent\textbf{Incorrect in the original:} The original results for domain
walls with diluted (fractional) excitons and with fixed atomic
coordinates are valid for a hypothetical ferroelectric that does
not exhibit carrier self-trapping. Such a material may or may
not exist. %We learn what a ferroelectric must be like in order
%to exhibit a large domain-wall photovoltage. 
For BiFeO$_3$ it
is necessary to include carrier self-trapping. The original results for
the delocalized exciton state should be valid for BiFeO$_3$ for
less than a picosecond after exciton formation (see below). The original
Figure~6, which shows the electronic potential without and
with a delocalized exciton and the domain-wall photovoltage
profile, is valid for a hypothetical ferroelectric without
self-trapping, or for BiFeO$_3$ before exciton self-trapping is
completed. Here in a modified Fig.~6 an additional panel 6(a) shows the densities
of excess electron and hole in the delocalized exciton state.
\\ \ \\\noindent
\textbf{Addition:} If we artificially fix the atomic coordinates in the presence of an electron-hole pair, corresponding
to an idealized ferroelectric without exciton self-trapping,
then the exciton is delocalized and split by the domain wall,
with the hole on the left and the electron on the right
hand side of the wall, i.e. the exciton has a dipole moment
antiparallel to the net polarization [Figs.~3(a) and 6(a)]. The potential change induced by the exciton
determines the domain-wall photovoltage except for the ionic
screening, which is added afterwards by means of an effective
dielectric constant.

%
%%%%%%%%%%%%%%%%%%%%%%%%%%%%%%%%%%%%%%%%%%%%%%%%%%%%%%%%%%%%%%%%%%%%%%%%%%%%%%%%%%%%%%%%%%%%%%%%%%%%%%%%
\subsubsection{Atomic coordinates optimized}
%%%%%%%%%%%%%%%%%%%%%%%%%%%%%%%%%%%%%%%%%%%%%%%%%%%%%%%%%%%%%%%%%%%%%%%%%%%%%%%%%%%%%%%%%%%%%%%%%%%%%%%%
\noindent\textbf{Incorrect in the original:} The original results for domain walls
with a fraction of an exciton and with optimized atomic
coordinates are wrong. The original claim that excitons are
delocalized below a critical exciton density is incorrect. The
top panel of the original Figure 3 showing a delocalized
spatial distribution of diluted excitons is wrong. The original
Figure 4, depicting densities of diluted excitons and exciton
formation energies after optimization of atomic coordinates,
is wrong. The original Figure 5, depicting a phase diagram
for excitons with stability regions of a delocalized and a
self-trapped exciton state, is wrong. Original text sections
corresponding to the original Figures 3(a), 4, and 5 are wrong.
\\ \ \\\noindent
\textbf{Correct in the original:} The original results for do-
main walls with an integer exciton in the supercell, depicted
in the original Fig.~3(b), are correct. %The original claim that
%excitons become self-trapped at high densities is correct for
%all considered densities.
\\ \ \\\noindent
\textbf{Addition:} Initially the exciton is split by the domain
wall and exhibits a dipole moment [see Fig.~3(a)]. When changes in the
atomic positions are allowed, the exciton becomes spontaneously self-trapped, 
see Fig.~\ref{fig:densities_small_large_X}(b). After self-trapping both
electron and hole reside on the same side of the domain wall
and the excitons's dipole moment is lost. Carrier self-trapping
destroys the domain-wall photovoltage.

\setcounter{figure}{2} 

\begin{figure}[htb]
  %\begin{tikzpicture}
  %%\draw (2,2) circle (3cm);
  %%(-4,4) node[shift={(-1.0,0.3)}]{$(a)$};
  %\fill[fill=green!40!black] (-1,2) ellipse (1cm and 1cm) node {\textcolor{white}{$h^+$}};
  %\fill[fill=red!40!orange] (1,2) ellipse (1cm and 1cm) node {\textcolor{white}{$e^-$}};
  %\draw[thick] (0,1) -- (0,3) node[anchor=south]{DW};
  %%\draw (4,2) circle (3cm);
  %\draw[thick,->] (2.5,2) -- (3.5,2) node[anchor=north, shift={(-0.4,0.0)}] {self-trapping}; % arrow
  %\fill[fill=green!40!black] (5,2) ellipse (0.75cm and 0.75cm) node[anchor=south,shift={(0.1,0.25)}] {\textcolor{white}{$h^+$}};
  %\fill[fill=red!40!orange] (5,2) ellipse (0.25cm and 0.25cm) node {\textcolor{white}{$e^-$}};
  %\draw[thick] (4.25,1) -- (4.25,3) node[anchor=south]{DW};
  %\end{tikzpicture}
  \begin{tikzpicture}
  %\draw (2,2) circle (3cm);
  \draw[thick] (-2,3.5) -- (-2,3.5) node[anchor=north, shift={(0.0,0.0)}] {(a)};
  \draw[thick] (3,3.5) -- (3,3.5) node[anchor=north, shift={(0.0,0.0)}] {(b)};
  \draw[thick] (-2,1) -- (-2,1) node[anchor=north, shift={(0.0,-0.25)}] {(c)};
  %(-4,4) node[shift={(-1.0,0.3)}]{$(a)$};
  \shade[right color=green!40!black,left color=white] (-2,1) rectangle (0,3) node[shift={(-0.5,-0.25)}] {\textcolor{white}{$h^+$}};
  \shade[left color=red!40!orange,right color=white] (0,1) rectangle (2,3) node[shift={(-1.5,-0.25)}] {\textcolor{white}{$e^-$}};
  \draw[thick] (0,1) -- (0,3) node[anchor=south]{DW};
  \draw[thick,<-] (0.5,1.9) -- (-0.5,1.9) node[anchor=north, shift={(0.225,0.45)}] {$P_{\mathrm{net}}$};
  \draw[thick,->] (0.5,1.5) -- (-0.5,1.5) node[anchor=north, shift={(0.25,0.05)}] {large dipole moment};
  \draw[thick] (0.5,1.5) -- (0.5,1.5) node[anchor=north, shift={(-0.2,-0.5)}] {instantaneous};
  %\draw (4,2) circle (3cm);
  \draw[thick,->] (2.5,2) -- (3.5,2) node[anchor=north, shift={(-0.4,0.0)}] {self-trapping}; % arrow
  \shade[inner color=green!40!black,outer color=white] (5,2) ellipse (0.75cm and 0.75cm) node[anchor=south,shift={(0.1,0.25)}] {\textcolor{white}{$h^+$}};
  \fill[inner color=red!40!orange,outer color=red!20!orange!30!white] (5,2) ellipse (0.25cm and 0.25cm) node {\textcolor{white}{$e^-$}};
  \draw[thick] (4.25,1) -- (4.25,3) node[anchor=south]{DW};
  \draw[thick] (4.5,1.5) -- (4.5,1.5) node[anchor=north, shift={(0.15,0.0)}] {small/no dipole moment};
  \draw[thick] (4.5,1.5) -- (4.5,1.5) node[anchor=north, shift={(-0.2,-0.5)}] {relaxed};
  \end{tikzpicture}
  \includegraphics[width=0.475\textwidth]{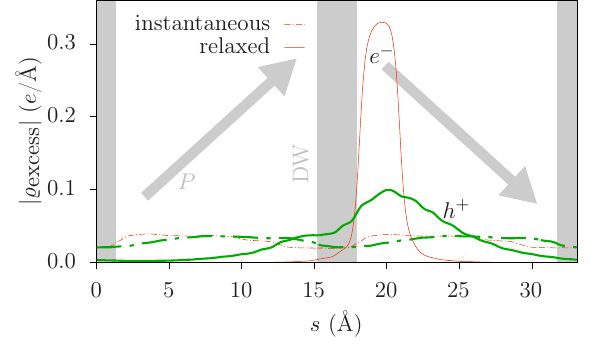}
  \caption{\label{fig:densities_small_large_X} \textbf{Modification:} (a) and (b) Sketch of self-trapping of an exciton at a ferroelectric domain wall (DW).
  (c)  Smoothened densities of excess electron and hole for 
	excitons at the 71\textdegree~domain wall before (``instantaneous'') and after (``relaxed'') 
optimization of the atomic coordinates, i. e. exciton self-trapping. The exciton density is 1~exciton per 120-atom supercell
  ($\approx2.3\cdot10^{14}$/cm$^2$).
	Gray bars and arrows indicate ferroelectric domains.
  }
\end{figure}

%%%%%%%%%%%%%%%%%%%%%%%%%%%%%%%%%%%%%%
\subsubsection{Ab initio molecular dynamics}
%%%%%%%%%%%%%%%%%%%%%%%%%%%%%%%%%%%%%%

\noindent\textbf{Addition:} How long does self-trapping of excitons take? By
means of molecular dynamics we show that the exciton self-trapping happens so fast that most of its lifetime the exciton is in the self-trapped state for
carrier lifetimes~$\gg$1~ps, see below. 
Figure~\ref{fig:MD_0K} shows the total energy change with respect to the delocalized exciton state, local
magnetic moments of Fe atoms, and real simulation temperature for a target temperature of 0~K as a function of time, and
the charge densities of excess electron and hole at different
times during the simulation. The energy gained by exciton self-trapping (small exciton polaron formation) is about
0.3~eV (the energy difference between $t = 0$ and $t\to\infty$).
Carrier self-trapping is revealed by the change in magnetic
moment of one Fe atom, where the excess electron localizes.
This localization happens abruptly in a time interval of the order of 10~fs [see inset in Fig.~\ref{fig:MD_0K}(b)]; it is accompanied by an
abrupt decrease in total energy [see inset in Fig.~\ref{fig:MD_0K}(a)]. Afterwards the system continuously lowers its energy further during several hundred femtoseconds [Fig. 4(a)]. No activation
barrier for localization is visible. Note that the precise numerical value of the local magnetic moment is not meaningful,
as it depends on the radius chosen for the projection on the
Fe atoms; the expected value for an Fe$^{3+}$ ion without excess
charge would be 5~$\mu_{\mathrm{B}}$ (observed here: $\approx 4~\mu_{\mathrm{B}}$).
At~300 K (Fig.~\ref{fig:MD_300K}) the exciton localizes instantaneously (one magnetic moment is much smaller than the others already in the first time step). 
Other than at 0~K, the exciton
localizes in some distance from the domain wall, indicating
that thermal fluctuations of atomic positions are more important than the trapping potential of the domain walls. 
The position of the self-trapped exciton changes between 0 and 9~fs,
indicating that the thermal energy at room temperature is sufficient to overcome the barrier for exciton hopping.
The second reduced magnetic moment belongs to an Fe atom
carrying a localized hole, see Fig.~\ref{fig:MD_300K}(b) and (f). We find that
the hole part of the exciton is strongly localized, other than
the individual hole \cite{koerbel:2018:electron}.
The time needed for carrier self-trapping (small polaron
formation) could increase with decreasing exciton density. Allowing for an unphysical influence of supercell size, thermostat parameter, and exciton density on the simulated dynamics, 
we estimate 1~ps as an upper limit for the time needed for
exciton self-trapping.

\begin{figure}[htb]
  \includegraphics[width=0.475\textwidth]{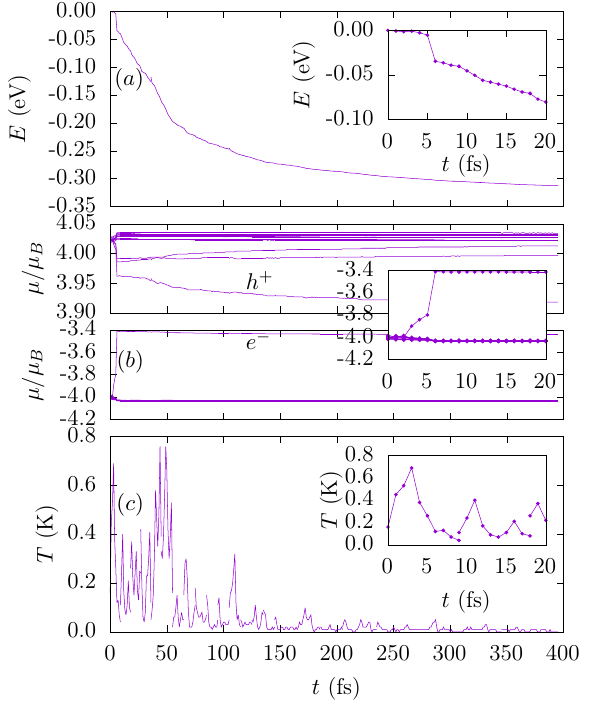}
  \includegraphics[width=0.39\textwidth]{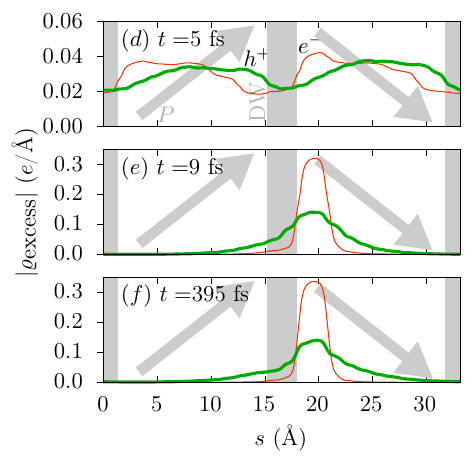}
	\caption{\label{fig:MD_0K} \textbf{Addition:} (a) Total energy change with respect to the delocalized exciton state, (b) local magnetic moments of Fe atoms, 
	and (c) temperature for an MD simulation with an electron-hole pair at a target temperature of 0~K.
	(d) to (f): planar averaged charge density of electron and hole at different times $t$.
  }
\end{figure}
\begin{figure}[htb]
  \includegraphics[width=0.475\textwidth]{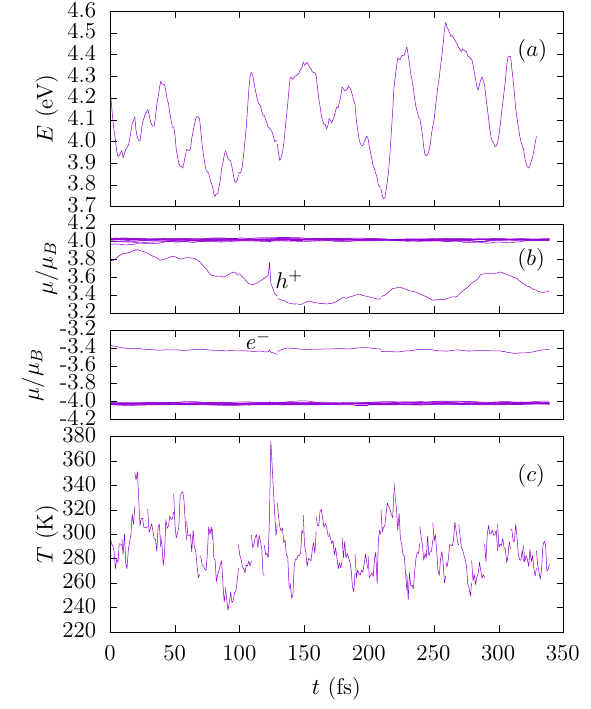}
  \includegraphics[width=0.39\textwidth]{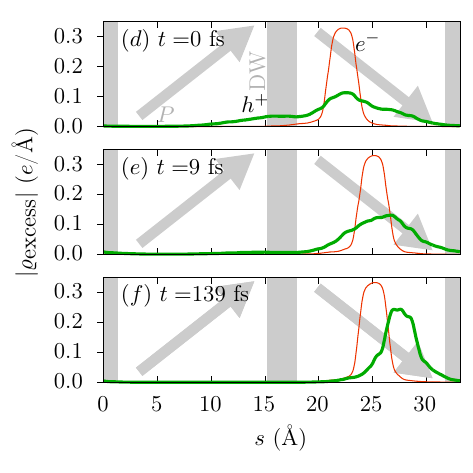}
	\caption{\label{fig:MD_300K} \textbf{Addition:} The same as Fig.~\ref{fig:MD_0K} for a target temperature of 300~K.
  }
\end{figure}

%%%%%%%%%%%%%%%%%%%%%%%%%%%%%%%%%%%%%%%%%%%%%%%%%%%%%%%%%%%%%%%%%%%%%%%%%%%%%%%%%%%%%%%%%%%%%%%%%%%%%%%%
\subsection{Domain-wall photovoltage}
%%%%%%%%%%%%%%%%%%%%%%%%%%%%%%%%%%%%%%%%%%%%%%%%%%%%%%%%%%%%%%%%%%%%%%%%%%%%%%%%%%%%%%%%%%%%%%%%%%%%%%%%
 \begin{figure}[htb]
  \includegraphics[width=0.5\textwidth]{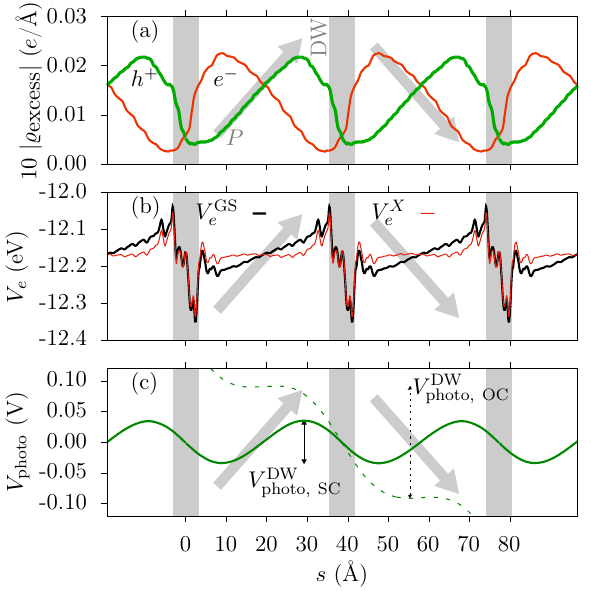}
  \caption{\label{fig:potential} 
	\textbf{Modification:} (a) Electron and hole density at the 71\textdegree~domain wall in the excited state 
     in a 280-atom supercell with 0.1 excitons.
    (b) Smoothened potential energy of electrons in the ground state ($V_e^{\mathrm{GS}}$)
    and in the excited state ($V_e^{X}$).
    (c) Instantaneous domain-wall photovoltage profile $V_{\mathrm{photo}}$ %[cf. Eq.~\eqref{eq:PV_profile}]
    [cf. Eq.~1]
    for short-circuit (solid line) and open-circuit conditions (dashed line).
	 Green arrows show the magnitude of the domain-wall photovoltage $V_{\mathrm{photo}}^{\mathrm{DW}}$.
  }
\end{figure}
\begin{figure*}[t]
\includegraphics[width=\textwidth]{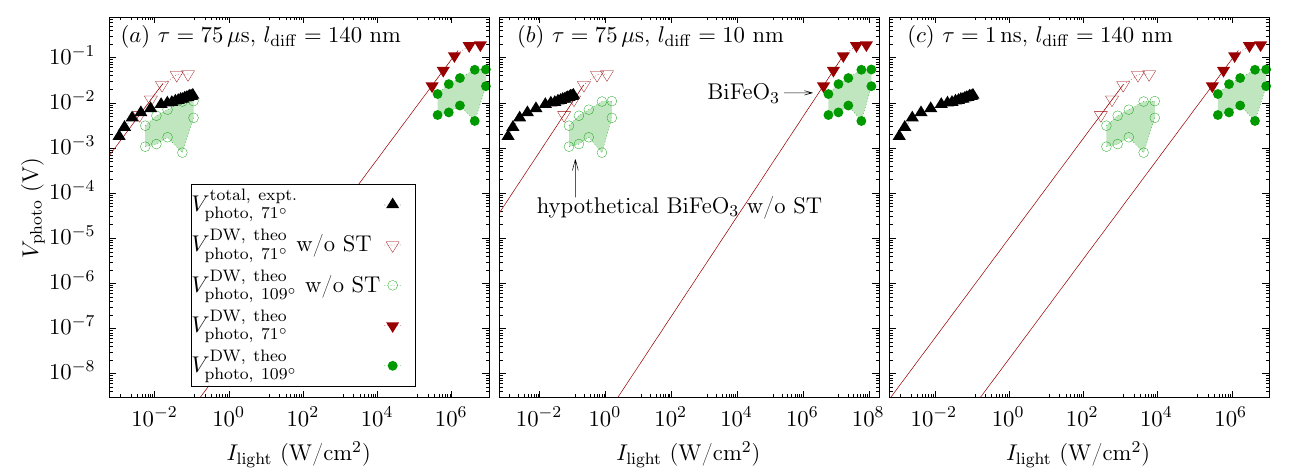}
\caption{\label{fig:U_I_light} 
  \textbf{Modification:} Measured photovoltage per domain of BiFeO$_3$ films with 71\textdegree~domain walls from Ref.~\onlinecite{seidel:2011:efficient} 
    and calculated domain-wall photovoltage of the 71\textdegree\ and the 109\textdegree\ domain walls  as a function of the illumination intensity
    for different photocarrier life times $\tau$ and diffusion lengths $l_{\mathrm{diff}}$.
	For a hypothetical BiFeO$_3$-like ferroelectric without self-trapping (w/o ST, empty symbols), the domain-wall photovoltage can reach the measured total photovoltage at similar light intensities as in experiment [(a) and (b)]; 
	when exciton self-trapping is taken into account (filled green and brown symbols), the domain-wall photovoltage is reduced by several orders of magnitude.
    Solid lines are an extrapolation, dotted lines are a guide to the eye. 
}
\end{figure*}

\noindent\textbf{Incorrect in the original:} The original Figure~7 showed the
domain-wall photovoltage of a hypothetical ferroelectric
without exciton self-trapping, not of BiFeO$_3$.
\\ \ \\\noindent
\textbf{Correction:} We consider two cases, one hypothetical,
one real. The hypothetical case is a BiFeO$_3$-like ferroelectric
without carrier self-trapping. Its domain-wall photovoltage is
calculated assuming that the excitons remain in their delocalized state and are split by the domain wall throughout their
lifetime. Figure~\ref{fig:U_I_light} depicts these hypothetical domain-wall
photovoltages (empty symbols), which are identical to those
in the original Fig.~7. In addition, the new Fig.~\ref{fig:U_I_light} depicts
also the domain-wall photovoltages of real BiFeO$_3$, which
are extremely small due to fast self-trapping of carriers, as
we saw in the molecular dynamics section. The self-trapped
exciton is no longer split over the two sides of the domain
wall, it has no considerable dipole moment. The domain-wall
photovoltages for this real case are calculated by assuming
that the domain-wall photovoltage is that of hypothetical
BiFeO$_3$ during self-trapping and zero after self-trapping (after
1~ps). This is an optimistic assumption. Figure~\ref{fig:U_I_light} shows the
time average of the calculated domain-wall photovoltage.
\indent For a hypothetical BiFeO$_3$-like ferroelectric
without self-trapping of excitons, the calculated domain-wall
photovoltage matches the experimentally measured total photovoltage per domain only in the most optimistic scenario
[Fig. 7(a)], in which we assume 
a carrier lifetime of 75~$\mu$s and a carrier diffusion length of 140~nm, 
resulting in a photocarrier density of $10^{16}$/cm$^3$ to $10^{18}$/cm$^3$.
For comparison, in Ref.~\cite{seidel:2011:efficient} a carrier density of about 10$^{12}$/cm$^3$ to $10^{13}$/cm$^3$, a carrier diffusion length of 8~nm, and a lifetime of 35~ps were assumed,
in Ref.~\onlinecite{sheu:2012:ultrafast} a lifetime of 1~ns. If we assume such conditions
[Fig. 7(b), (c)], the DW-PVE is orders of magnitude too small
to account for the major part of the measured photovoltage
in BiFeO$_3$ even when carrier self-trapping does not occur. In real BiFeO$_3$, where we observe spontaneous
self-trapping of excitons on a sub-ps time scale, the domain-wall photovoltage is orders of magnitude lower than that of
the hypothetical material and the measured total photovoltage,
which shows that pristine ferroelectric domain walls cannot be
responsible for the measured photovoltage of BiFeO$_3$.

%%%%%%%%%%%%%%%%%%%%%%%%%%%%%%%%%%%%%%%%%%%%%%%%%%%%%%%%%%%%%%%%%%%%%%%%%%%%%%%%%%%%%%%%%%%%%%%%%%%%%%%%
\section{Conclusion} 
%%%%%%%%%%%%%%%%%%%%%%%%%%%%%%%%%%%%%%%%%%%%%%%%%%%%%%%%%%%%%%%%%%%%%%%%%%%%%%%%%%%%%%%%%%%%%%%%%%%%%%%%
\noindent\textbf{Correct in the original:} The original conclusions that
``the ferroelectric polarization profile does not allow one to
determine the correct sign and magnitude of the electrostatic
potential at the domain walls'' and that the self-trapped
exciton could be experimentally detected ``as a state inside the
band gap in photoluminescence spectroscopy'' are correct.
\\ \ \\\noindent
\textbf{Incorrect in the original:} The original conclusion that the
domain-wall photovoltage in BiFeO$_3$ might be responsible
for the measured photovoltage is wrong.
\\ \ \\\noindent
\textbf{Correction}: 
%We have analyzed the contribution of ferroelectric domain walls to the
%photovoltage in BiFeO$_3$ using first principles methods. We
%find that the ferroelectric polarization profile does not allow to
%determine the correct sign and magnitude of the electrostatic
%potential at the domain walls. Instead the electronic potential
%should be directly determined from \textit{ab initio} calculations.
\indent
In principle, the domain-wall driven photovoltage in a
ferroelectric can be as large as $\approx$10~mV per domain wall in
ferroelectrics that do not exhibit exciton self-trapping, if the
carrier lifetimes and carrier diffusion lengths are of the order
of $\approx$100~$\mu$s and $\approx$100~nm, respectively.
\\\indent
We find that this is not the case for BiFeO$_3$. We observe
strong exciton self-trapping on a sub-ps timescale, which
overcomes charge separation by the ferroelectric domain
walls and renders the domain-wall driven photovoltage several orders of magnitude lower than the total experimentally
measured photovoltage of $\approx$10~mV per 140-nm domain.
Hence, the measured photovoltage in BiFeO$_3$ must arise
from causes other than pristine ferroelectric domain walls.
Note that here only domain-wall photovoltaic effects were
considered, not bulk effects.
%\\\indent
%The self-trapped exciton could be experimentally detected
%by means of photoluminescence spectroscopy as a state with
%an energy level inside the bandgap.

%
\section*{Acknowledgement}
This project  has  received  funding  from  the  
European Union's Horizon 2020 research and innovation programme under the Marie Sk\l{}odowska-Curie 
Grant Agreement No. 746964 - FERROVOLT.
Computation time and support provided by 
the Trinity Centre for High Performance Computing funded by Science Foundation Ireland is gratefully acknowledged. 
We also thank the Irish Centre for High-End Computing (ICHEC) for the provision of computational facilities and support.
%Further computational resources were supplied by the project "e-Infrastruktura CZ" (e-INFRA LM2018140) provided within the program Projects of Large Research, Development and Innovations Infrastructures.
We are grateful to Subhayan Roychoudhury (Trinity College Dublin) for helpful discussions about excited-state density-functional theory.
%and for pointing out relevant literature.
Graphics were made using gnuplot 
 and VESTA \cite{momma:2011:vesta}.
%
%\nocite{moreau:1971:ferroelectric}
%\nocite{palewicz:2007:atomic}    
%\nocite{palewicz:2010:bifeo3}    
%\nocite{fischer:1980:temperature}
%\nocite{lebeugle:2007:very}      

%% {moreau:1971:ferroelectric}%
%% {palewicz:2007:atomic}%
%% {palewicz:2010:bifeo3}%
%% {fischer:1980:temperature}%
%% {lebeugle:2007:very}%
%% {wang:2013:bifeo3}%
%% {dieguez:2013:domain}%
%% {chen:2017:polar}%
%% {ren:2013:ferroelectric}%
%% {meyer:2002:ab}%
%% {zelezny:2010:optical}%

%\nocite{alexe:2011:tip}
%\nocite{schleich:2001:quantum}%

\bibliography{jabbr,all}

%%%%%%%%%%%%%%%%%%
%\newpage
%\section*{Supp. Inf.}
%%%%%%%%%%%%%%%%%%

\end{document}

\title{\reallynew{Erratum:} Photovoltage from ferroelectric domain walls in BiFeO$_3$:\\
\reallynew{Revised} Supplemental Material}

\author{Sabine K\"{o}rbel}
\affiliation{\dublin}
\email{skoerbel@uni-muenster.de}
\author{Stefano Sanvito}
\affiliation{\dublin}

\begin{abstract}
	\noindent \reallynew{Figure 4(a) in the original Supplemental Material was incorrect. So was part of the section ``Excitonic densities''.
	In this revised Supplemental Material, errors have been corrected and a new Figure 10 and a new section ``Self-trapping at low carrier concentrations'' have been added. 
	}
\end{abstract}
\maketitle%

%%%%%%%%%%%%%%%%%%%%%%%%%%%%%%%%%%%%%%%%%%%%%%%%%%%%%%%%%%%%%%%
\subsection*{Calculated ground-state properties of BiFeO$_3$}
\begin{table}[htb]
      \begin{tabular}{l | l | c | c  }
        \ \ \ \ \ \ \ \ \ &                                                                                                 & \new{ideal}               &                     \\
        \ \ \ \ \ \ \ \ \ &   Expt                                                                                          & \new{perovskite}          &  this work          \\\hline
        $a$ (\AA)         & 5.57330(5)\footnotemark[4]; 5.57882(5)\footnotemark[2];                                         &                           & 5.5006              \\
        \ \ \ \ \ \ \ \ \ & 5.5876(3)\footnotemark[1]; 5.58132(5)\footnotemark[3];                                          &                           & (-1.3\%)            \\\hline
        $c$  (\AA)        & 13.84238(16)\footnotemark[4]; 13.867(1)\footnotemark[1];                                        &                           & 13.507              \\
        \ \ \ \ \ \ \ \ \ & 13.86932(16)\footnotemark[2]; 13.87698(15)\footnotemark[3];                                     &                           & (-2.4\%)            \\\hline
        $c/a$             & 2.4817\footnotemark[1]; 2.48370\footnotemark[4];                                                & $\sqrt{6}$                & 2.4556              \\
        \ \ \ \ \ \ \ \ \ &  2.48607\footnotemark[2]; 2.48633\footnotemark[3]                                               & $\approx$2.4495           & (-81\%)             \\\hline
        $V$ (\AA$^3$)     & 372.36\footnotemark[4]; 373.06\footnotemark[6]; 373.83\footnotemark[2];                         &                           & 342.75              \\
                          & 374.37\footnotemark[3]; 374.94\footnotemark[1]; 375.05\footnotemark[5]                          &                           & (-8\%)              \\\hline
        $z_{\rm Fe}$      &  0.22021(09)\footnotemark[2]; 0.22046(8)\footnotemark[4];                                       & 0.25                      & 0.227               \\ 
        \ \ \ \ \ \ \ \ \ & 0.22067(8)\footnotemark[3];  0.2209(5)\footnotemark[5];                                         &                           & (-20\%)             \\
        \ \ \ \ \ \ \ \ \ & 0.2209(6)\footnotemark[6]; 0.2212(15)\footnotemark[1];                                          &                           &                     \\\hline
        $x_{\rm O}$       & 0.443(2)\footnotemark[1];  0.44506(22)\footnotemark[4];                                         & 0.5                       & 0.436               \\
        \ \ \ \ \ \ \ \ \ & 0.4453(8)\footnotemark[6]; 0.44582(22)\footnotemark[3];                                         &                           & (+12\%)             \\
        \ \ \ \ \ \ \ \ \ & 0.4460(8)\footnotemark[5] ;  0.44694(28)\footnotemark[2]                                        &                           &                     \\\hline
        $y_{\rm O}$       & 0.012(4)\footnotemark[1]; 0.01700(28)\footnotemark[3];                                          & 0.0                       & 0.018               \\
        \ \ \ \ \ \ \ \ \ &  0.0173(12)\footnotemark[5]; 0.01789(28)\footnotemark[4];                                       &                           & ($\pm$0\%)          \\
        \ \ \ \ \ \ \ \ \ & 0.01814(35)\footnotemark[2]; 0.0185(13)\footnotemark[6]                                         &                           &                     \\\hline
        $z_{\rm O}$       & 0.9511(5)\footnotemark[6]; 0.9513(5)\footnotemark[5];                                           & 0.0                       & 0.961               \\
        \ \ \ \ \ \ \ \ \ & 0.95152(11)\footnotemark[4] 0.95183(14)\footnotemark[2];                                        &                           & (-15\%)             \\
        \ \ \ \ \ \ \ \ \ & 0.95183(11)\footnotemark[3];  0.9543(20)\footnotemark[1]                                        &                           &                     \\\hline
 $\alpha$ (\textdegree)   & 59.34\footnotemark[3]; 59.35\footnotemark[2]; 59.35\footnotemark[5];                            & 60                        & 59.89               \\
        \ \ \ \ \ \ \ \ \ &  59.39\footnotemark[4]; 59.39\footnotemark[6]; 59.42\footnotemark[1]                            &                           & (-81\%)             \\\hline
  $\omega$ (\textdegree)  & 12.2\footnotemark[5]; 12.3\footnotemark[2]; 12.4\footnotemark[3];                               & 0                         & 14.4                \\
                          & 12.5\footnotemark[1]; 12.5\footnotemark[6];  12.6\footnotemark[4]                               &                           & (+14\%)             \\\hline
        $P$               & $\approx 100$\footnotemark[7];                                                                  &                           & 94                  \\ 
      \end{tabular}
      \footnotetext[1]{Moreau {\em et al.}   \cite{moreau:1971:ferroelectric}. Single crystal X-Ray and neutron powder diffraction at room temperature.}
      \footnotetext[2]{Palewicz {\em et al.} \cite{palewicz:2007:atomic}.     Neutron powder diffraction at 298~K.}
      \footnotetext[3]{Palewicz {\em et al.} \cite{palewicz:2010:bifeo3}.     Neutron powder diffraction at 298~K.}
      \footnotetext[4]{Palewicz {\em et al.} \cite{palewicz:2010:bifeo3}.     Neutron powder diffraction at 5~K.}
      \footnotetext[5]{Fischer {\em et al.}  \cite{fischer:1980:temperature}. Neutron powder diffraction at 293~K.}
      \footnotetext[6]{Fischer {\em et al.}  \cite{fischer:1980:temperature}. Neutron powder diffraction at 4.2~K.}
      \footnotetext[7]{Lebeugle {\em et al.} \cite{lebeugle:2007:very}.       Single crystals, $R3c$ phase, room temperature.}
      \caption{\label{tab:BFO_GS_props}\new{Measured (ordered by size) and calculated structure parameters of BiFeO$_3$ in the $R3c$ phase: 
        Hexagonal lattice constants $a$ and $c$, $c/a$ ratio, cell volume $V$, 
        fractional atomic coordinates, rhombohedral cell angle $\alpha$, octahedral rotation angle $\omega$, and ferroelectric polarization $P$.
        Errors (in brackets) are with respect to the closest experimental value.
        Errors of atomic coordinates, $\alpha$, and $c/a$ are those of ferroelectric distortions from the ideal perovskite structure}.
     }
\end{table}
%%\todo{add polarization ?. Done}
%\todo{add numbers for increased volume ? }

\new{Table~\ref{tab:BFO_GS_props} contains our calculated structural parameters of BiFeO$_3$ in the $R3c$ phase in comparison with experimental data.
Lattice parameters deviate from experiment by about 2\%, typical for LDA calculations.
Ferroelectric displacements of atoms from the ideal perovskite structure deviate by up to 20\%.
Ferroelectric distortions of the unit cell ($c/a$ ratio and rhombohedral cell angle) are severely underestimated by $\approx 80\%$,
nevertheless the ferroelectric polarization is in good agreement with experiment, indicating that ferroelectric properties are 
well captured by our computational setup. 
}
%\todo{rm table above, include Palewicz (two papers at least, done), Bucci, and others maybe. Remove errors ? Or redefine them. Done}

%%%%%%%%%%%%%%%%%%%%%%%%%%%%%%%%%%%%%%%%%%%%%%%%%%%%%%%%%%%%%%%
\subsection*{Calculated domain-wall properties}
%%%%%%%%%%%%%%%%%%%%%%%%%%%%%%%%%%%%%%%%%%%%%%%%%%%%%%%%%%%%%%%

In Table~\ref{tab:DW_properties} we compare our calculated DW formation energies and widths for different supercell sizes with those available in the literature.
For historical reasons, we include the 180\textdegree\ wall, although we do not consider it in the main article.
Note that Ref.~\onlinecite{wang:2013:bifeo3} found close agreement between domain-wall structures seen in electron miscroscopy and those calculated from first principles.
This is \new{true in particular} for the domain-wall widths, which \new{are as narrow as} about one atomic layer in the case of the 109\textdegree\ domain wall
and two atomic layers in the case of the 180\textdegree\ domain wall, both \new{according to experiment and to} first-principles calculations 
(see Fig.~1 and 2 in Ref.~\onlinecite{wang:2013:bifeo3}).
The 71\textdegree\ domain wall is slightly broader and extends over about three atomic planes. 
We fitted the polarization and tilt profiles $P$ and $A$ with a tangens hyperbolicus to extract the DW width $\xi_P$,
\begin{equation}
  \label{eq:pol_profile}
  P_r(s) = P_r^{\infty}\tanh[(s - s_0)/\xi_P]\:,
\end{equation}
where $P_r$ is the rotating component of the polarization, $P_r^{\infty}$ is its asymptotic value far away from the domain wall, and $s$ is the distance from the domain wall.
%We use a coordinate system with axes $\vect{e}_r$, $\vect{e}_s$, $\vect{e}_t$, where $\vect{e}_r\parallel P_r$, $\vect{e}_s$ is perpendicular to the domain-wall plane, and $\vect{e}_t=\vect{e}_r\times\vect{e}_s$.
\new{In this section, different from below and the main article, polarization profiles are calculated from ionic positions} and formal ionic charges (Bi$^{3+}$, Fe$^{3+}$, and O$^{2-}$).
Tests indicated that applying the more sophisticated Born effective charges changed the resulting polarization essentially only by a prefactor.
%We therefore expect only small differences between profiles from formal and Born-effective charges.
%
     \begin{table}
      \begin{tabular}{l | c | c | c | c | c }
                   &                 &    \multicolumn{2}{c |}{$E$ (mJ/m$^2$)}                                                   &                 &                           \\               
%        type       & $n_{\rm atoms}$ & this work& Ref.~\onlinecite{dieguez:2013:domain} &Ref.~\onlinecite{wang:2013:bifeo3}&Ref.~\onlinecite{chen:2017:polar} &  $\xi_P$ (nm)   &      $\xi_A$ (nm)  \\\hline
        type       & $n_{\rm atoms}$ & this work& Lit.                                         &        $\xi_P$ (\AA)   &      $\xi_A$ (\AA)  \\\hline
        71\degree  & 80              & 167      &   152\footnotemark[1], 156\footnotemark[4]   & 2.94$\pm$0.01   &   3.4$\pm$0.7      \\
        71\degree  &100              & 170      &                                              & 2.94$\pm$0.05   &   2.4$\pm$0.4       \\
        71\degree  &120              & 172      &   167\footnotemark[1], 143.2\footnotemark[3] & 2.94$\pm$0.04   &   2.6$\pm$0.3         \\
        71\degree  &160              &          &                        128\footnotemark[2]   &                 &                        \\\hline
        109\degree & 80              & 60       &   62\footnotemark[1], 53\footnotemark[4]     & 1.8$\pm$0.2     &   0.4 \\ %$\pm$10$^6$            \\
        109\degree &100              & 60       &                                              & 2.0$\pm$0.1     &   0.4 \\ %$\pm$1045             \\
        109\degree &120              & 63       &   62\footnotemark[1], 52.9\footnotemark[3]   & 1.9$\pm$0.1     &   0.1 \\ %$\pm 6\cdot 10^{25}$    \\
        109\degree &160              &          &                        33\footnotemark[2]    &                 &                           \\\hline
        180\degree & 80              & 94       &   74\footnotemark[1], 71\footnotemark[4]                        & 1.63$\pm$0.06   &   --                \\
        180\degree &100              & 86       &                                              & 1.67$\pm$0.05   &   --                \\
        180\degree &120              & 84       &   82\footnotemark[1], 81.9\footnotemark[3]   & 1.67$\pm$0.04   &   --                \\
        180\degree &140              & 86       &                                              & 1.67$\pm$0.04   &   --              \\
        180\degree &160              &          &                        98\footnotemark[2]    &                 &                               
      \end{tabular}
      \footnotetext[1]{Di\'{e}guez{ \it et al.}\cite{dieguez:2013:domain}, LDA$+U$, $U$=4~eV}
    \footnotetext[2]{Wang{ \it et al.}\cite{wang:2013:bifeo3}, GGA$+U$, $U$=7~eV, $J$=1~eV}
    \footnotetext[3]{Chen{ \it et al.}\cite{chen:2017:polar}, GGA$+U$, $U$=3~eV}
    \footnotetext[4]{Ren{ \it et al.}\cite{ren:2013:ferroelectric}, LDA$+U$, $U$=3.87~eV}
      \caption{\label{tab:DW_properties}DW energies $E$ and DW widths ($\xi_P$: polarization, $\xi_A$: tilt wall width) as function of the supercell size. Errors are standard errors of the fit.}
     \end{table}
     \new{The DW energies} are close to those of Di\'{e}guez \cite{dieguez:2013:domain} and do not change very strongly with supercell size. 
     \new{Polarization and} tilt profiles (Fig.~\ref{fig:pol_profiles}) are also converged with respect to supercell size. 
%We finally chose supercells with 120 atoms to obtain the results presented in the main article.
%At this supercell size, the formation energy and the width of the domain walls are approximately converged.
% \new{The aim of this section is to give an impression of the convergence behavior of domain-wall properties in the ground state, without excitons,
% and to demonstrate that the domain walls are as narrow as a few atomic planes.}

 \begin{figure}[htb]
  \includegraphics[width=0.4\textwidth]{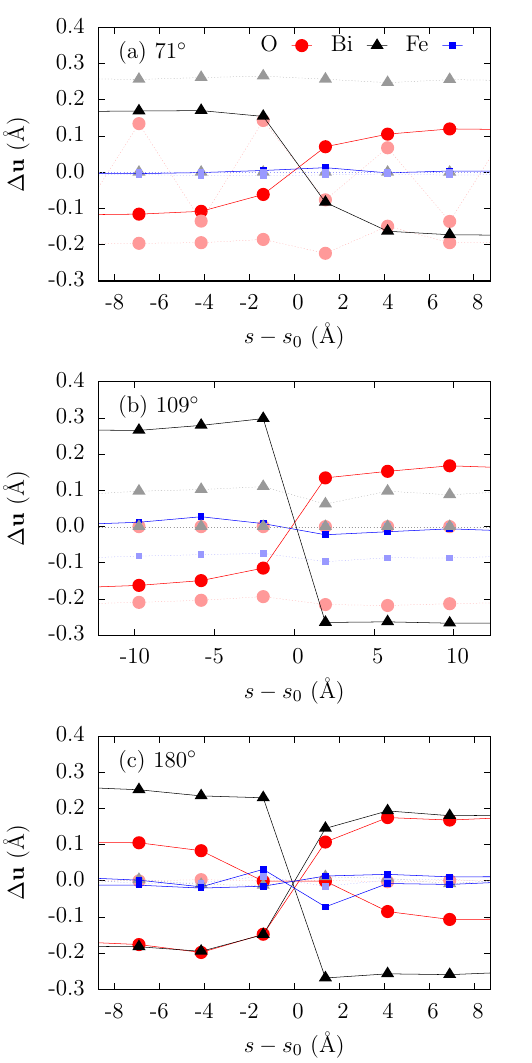}
  \caption{\label{fig:pos_profiles}\new{The three cartesian components of the atomic displacements from the paraelectric, tilt-free phase in each atomic layer 
    (a) for the 71\textdegree\ domain wall, (b) for the 109\textdegree\ domain wall, and (c) for the 180\textdegree\ domain wall in
      the coordinate system spanned by the supercell vectors: pseudocubic $[110]$, $[\bar{1}10]$, and $[001]$ directions (71\textdegree\ wall); 
      $[100]$, $[011]$, and $[0\bar{1}1]$ (109\textdegree\ wall); $[1\bar{1}0]$, $[110]$, and $[001]$ (180\textdegree\ wall).
  Components that nominally remain constant across the domain wall are transparent. $s_0$ is the position of the domain wall.
}}
\end{figure}

Figure~\ref{fig:pos_profiles} shows the layer-resolved ionic sublattice displacements compared to \new{a paraelectric reference structure without octahedral tilts}. 
The displacement profiles are atomically sharp. 
%
%\begin{figure*}[htbp]
%  \includegraphics[width=0.45\textwidth]{pol_profiles.pdf}
%    \includegraphics[width=0.45\textwidth]{tilt_profiles.pdf}
%    \caption{\label{fig:pol_profiles} (Color online) Polarization $P$ and tilt $A$ profiles for different supercell sizes (a) and (d) for the 71\textdegree, (b) and (e) for the 109\textdegree, and (c) and (f) for the 180\textdegree\ DW. Components that do not change sign at the DW are drawn transparent. Dashed transparent lines are the bulk reference. 
%  The profiles were fit according to Eq.~\ref{eq:pol_profile}. $s_0$ is the position of the domain wall.
%  }
%\end{figure*}
\begin{figure}[htbp]
  \includegraphics[width=0.45\textwidth]{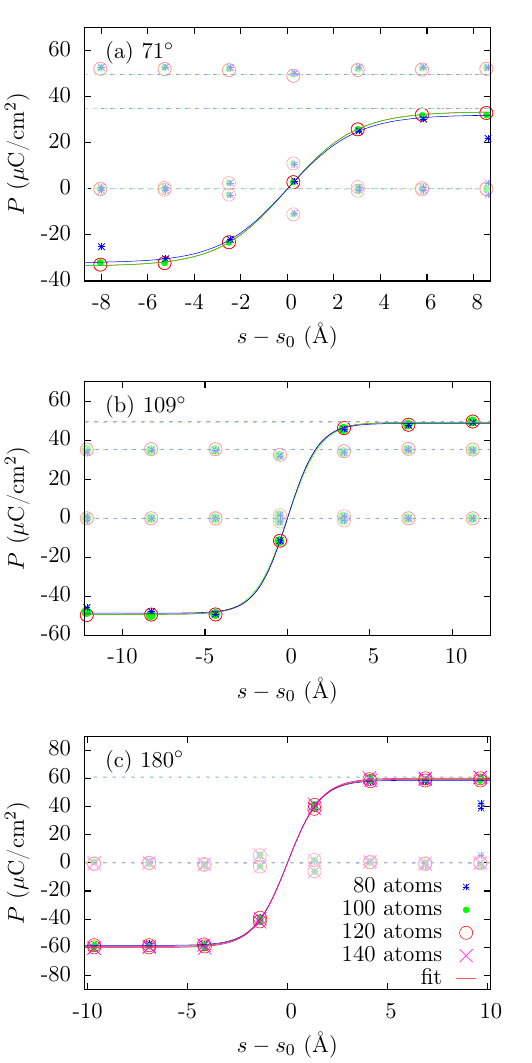}
    \caption{\label{fig:pol_profiles} Polarization $P$ profiles for different supercell sizes (a) for the 71\textdegree, (b) for the 109\textdegree domain wall,
      and (c) for the 180\textdegree\ wall.
      Components that do not change sign at the domain wall are drawn transparent. 
      Dashed transparent lines are the bulk reference. 
      The profiles were fit according to Eq.~\eqref{eq:pol_profile}. $s_0$ is the position of the domain wall.
  }
\end{figure}
%%%%%%%%%%%%%%%%%%%%%%%%%%%%%%%%%%%%%%%%%%%%%%%%%%%%%%%%%%%%%%%%%%%%%%%
%\subsection*{Electronic potential at the 109\textdegree~domain wall}
%%%%%%%%%%%%%%%%%%%%%%%%%%%%%%%%%%%%%%%%%%%%%%%%%%%%%%%%%%%%%%%%%%%%%%%
%\begin{figure}
%   \def\svgwidth{0.475\textwidth}
%   \input{DW_structure_109.pdf_tex}
%   \includegraphics[width=0.475\textwidth]{pol_pot_109DW_GS_240at.pdf}
%   \caption{\label{fig:GS}
%   %(a) Supercell consisting of 200 atoms with 109\textdegree\ domain walls,
%   %(b) three components of the ionic polarization, (c) ionic polarization component $P_s$ across the domain wall, 
%   %(d) electronic potential energy $V_e^{P_s}$ (thin blue lines) calculated from $P_s$, and the true electronic potential energy $V_e$ (thick black lines), shifted to zero. 
%   %The potential energy for open-circuit conditions is drawn with dashed lines.
%   The same as Fig.~1 in the main article, but for the 109\textdegree~domain walls, and with 240 atoms in the supercell.
% }
%\end{figure}
%Figure~\ref{fig:GS} shows the polarization and electronic potential of the 109\textdegree~domain wall in the ground state. 
%The real potential has a pronounced minimum at the domain wall, and a small slope in the domain interior, which yields a potential step at the wall of about 0.009~V. 
%If the ionic polarization is used instead to calculate an electrostatic potential, one obtains a grossly overrated potential step of about 0.2~V. 
%However, for the 109\textdegree~domain wall, other than for the 71\textdegree~domain wall, at least the sign of the potential step is correct in the polarization-based approach.

\new{
%%%%%%%%%%%%%%%%%%%%%%%%%%%%%%%%%%%%%%%%%%%%%%%%%%%%%%%%%%%%%%%%%%%%%%%
\subsection*{Electronic potential step at the domain wall}
%%%%%%%%%%%%%%%%%%%%%%%%%%%%%%%%%%%%%%%%%%%%%%%%%%%%%%%%%%%%%%%%%%%%%%%
Figure~\ref{fig:delta_V_DW_GS} depicts the magnitude of the electrostatic open-circuit potential step at the domain wall 
in the ground state (without excitons). These data are also listed in Table~\ref{tab:delta_V_DW_GS}.
\begin{table}[htb]
      \begin{tabular}{l |  c | c | c | c}
        &   \multicolumn{2}{c}{$-\Delta V_e^{\mathrm{DW}}$} (mV)    & \multicolumn{2}{|c}{$d_{\mathrm{DW}}$ (\AA)}\\  
        $n_{\mathrm{atoms}}$ &  71\textdegree    &  109\textdegree  &  71\textdegree    &  109\textdegree              \\\hline
        160     & 187&  -18.3                                       &22.1               & 31.2    \\
        200     & 159&   -5.8                                       &27.6               & 39.0    \\
        240     & 150&    2.9                                       &33.2               & 46.8    \\
        280     & 157&    5.6                                       &38.7               & 54.6    \\
        $\infty$& 133&   17.3                                       &$\infty$           &$\infty$ 
      \end{tabular}
      \caption{\label{tab:delta_V_DW_GS}
        Electrostatic open-circuit potential step $-\Delta V_e^{\mathrm{DW}}$ of the 71\textdegree~and 109\textdegree~domain walls without excitons 
        as a function of the domain-wall distance $d_{\mathrm{DW}}$ (the number of atoms in the supercell, $n_{\mathrm{atoms}}$), as depicted in Fig.~\ref{fig:delta_V_DW_GS}.
      }
\end{table}
The potential converges to a positive number for both walls, 
so that it has the opposite sign compared to the polarization-based potential. 
We \new{obtain} the limit $\Delta V_e^{\mathrm{DW}}\left(\infty\right)$ of the potential step for large domain-wall distances $d_{\mathrm{DW}}$ by fitting with a power law:
\begin{equation}\label{eq:delta_V_DW_GS_fit}
  \Delta V_e^{\mathrm{DW}}\left(d_{\mathrm{DW}}\right)=\Delta V_e^{\mathrm{DW}}\left(\infty\right)+c\cdot d_{\mathrm{DW}}^p,
\end{equation}
\noindent where $\Delta V_e^{\mathrm{DW}}\left(\infty\right)$, $c$, and $p$ are fit parameters. 
The resulting exponents are $p_{\mathrm{71\degree~DW}}=-2.00$ and $p_{\mathrm{109\degree~DW}}=-2.07$.
\begin{figure}[htbp]
  \includegraphics[width=0.4\textwidth]{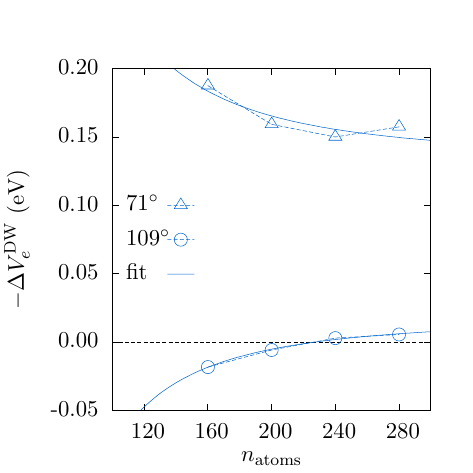}
    \caption{\label{fig:delta_V_DW_GS} Electrostatic potential step (electronic potential drop) 
      at the 71\textdegree~ and the 109\textdegree~ domain wall
      without excitons as a function of the number of atoms $n_{\mathrm{atoms}}$ in the supercell 
      (which is proportional to the domain-wall distance).
  }
\end{figure}
}
%%%%%%%%%%%%%%%%%%%%%%%%%%%%%%%%%%%%%%%%%%%%%%%%%%%%%%%%%%%%%%%%%%%%%%%
\subsection*{Calculation of the polarization-based potential}
%%%%%%%%%%%%%%%%%%%%%%%%%%%%%%%%%%%%%%%%%%%%%%%%%%%%%%%%%%%%%%%%%%%%%%%
\new{
  The ferroelectric polarization was calculated from ionic displacements from ideal perovskite positions using direction-averaged Born effective charges $Z^*$ of the $R3c$ phase 
  ($Z_{\mathrm{Bi}}^*=4.865$, $Z_{\mathrm{Fe}}^*=3.886$, $Z_{\mathrm{O}}^*=-2.917$).
  First a smooth curve was fitted to the polarization profile perpendicular to the domain wall [$P_s$ in Fig.~1~(c) and Fig.~2~(c) in the main article]
   using the program gnuplot. 
  We employed a symmetric fit function of the form
  \begin{equation}\label{eq:fit1}
    P_s(s)=P_s^{(0)}-\frac{P_s^{(1)}}{\cosh^2\left(\frac{s-s_0}{\xi}\right)},
  \end{equation}
  where $P_s^0$, $P_s^1$, $s_0$, and $\xi$ are fit parameters.
%
%  In the case of the 109\textdegree~wall we added an antisymmetric part to Eq.~\eqref{eq:fit1} to reduce the maximum point-wise deviation between fitted curve and data points:
%  \begin{equation}\label{eq:fit2}
%    P_s(s)=P_s^{(0)}-\frac{P_s^{(1)}}{\cosh^2\left(\frac{s-s_0}{\xi}\right)} + P_s^{(2)} \frac{\tanh\left(\frac{s-s_0}{\xi}\right)}{\cosh^2\left(\frac{s-s_0}{\xi}\right)}.
%  \end{equation}
%  \noindent $P_s^{(2)}$ was fixed to $P_s^{(2)}=4\ \mathrm{\mu C/m}^2$ to avoid large point-wise deviations between fitted curve and data points.
  The polarization-based electronic potential was then calculated as (compare Eq.~(8) in Ref.~\cite{meyer:2002:ab})
  \begin{equation}\label{eq:pot_hypo}
    V_e^{\mathrm{pb}}(s)= - \frac{1}{\varepsilon_{ss}}\int_{s^-}^s (P_s(s')-P_s(s^-)) \,\mathrm{d} s',
  \end{equation}
  \noindent where $s^-$ is a position in the domain interior on the left hand side of the domain wall, and $\varepsilon_{ss}$ is the calculated electronic dielectric constant for an electric field perpendicular to the domain-wall plane (see below).
  For $P_s(s)$ we used the fitted curve from Eq.~\eqref{eq:fit1}. % and Eq.~\eqref{eq:fit2}.
}
\new{
%%%%%%%%%%%%%%%%%%%%%%%%%%%%%
\subsection*{Electronic and lattice screening}
%%%%%%%%%%%%%%%%%%%%%%%%%%%%%
The screening (the real part of the static dielectric constant $\varepsilon$) is calculated {\em ab initio}
using the primitive rhombohedral cell and a $k$-point mesh of $10\times 10\times 10$ points.
$\varepsilon$ is diagonal in the coordinate system spanned by the hexagonal lattice vectors (the pseudocubic 
$[111]$, $[\bar{1}10]$, and $[\bar{1}\bar{1}2]$ directions, parallel and perpendicular to the ferroelectric polarization).
The electronic contribution $\varepsilon^e$ has the eigenvalues 7.4, 8.1, and 8.1,
 the lattice (ionic) contribution has the eigenvalues 21, 37, and 37.
In order to obtain the screening in the direction $s$ perpendicular to a domain wall, 
one needs to rotate $\varepsilon$ and obtains $\varepsilon^e_{ss}=7.6$, $\varepsilon^{\mathrm{ion}}_{ss}=26$, and $\varepsilon^{\mathrm{total}}_{ss}=34$ for the 71\textdegree\ domain wall 
and $\varepsilon^{e}_{ss}=7.9$, $\varepsilon^{\mathrm{ion}}_{ss}=32$, and $\varepsilon^{\mathrm{total}}_{ss}=40$ for the 109\textdegree\ domain wall.
Both in the case of the polarization-based potential in the ground state and in the case of the photovoltage, 
screening is included {\em a posteriori}. For the polarization-based potential we use $\varepsilon_{\mathrm{total}}^{ss}$,
for the photovoltages we use an effective 
dielectric constant $\varepsilon^{\mathrm{eff}}_{ss}$ that replaces the electronic screening $\varepsilon^e_{ss}$ that is automatically included by the total (electronic and lattice) screening:
\begin{equation}\label{eq:epsilon_eff}
  \varepsilon^{\mathrm{eff}}_{ss}=\frac{\varepsilon^{\mathrm{ion}}_{ss}+\varepsilon^e_{ss}}{\varepsilon^e_{ss}}\,.
\end{equation}
The effective dielectric constant is $\varepsilon_{ss}^{\mathrm{eff}}=4.4$ for the 71\textdegree\ DW and $\varepsilon_{ss}^{\mathrm{eff}}=5.0$ for the 109\textdegree\ DW.
}
%\todo{add screened potential in table, both GS and LEH. Done for LEH, GS does not make sense.}
%\todo{calculate and use epsilon along s for GS hypo pot. Done.}
\new{
%%%%%%%%%%%%%%%%%%%%%%%%%%%%%%
\subsection*{Excitonic densities}
%%%%%%%%%%%%%%%%%%%%%%%%%%%%%%
Figure~\ref{fig:densities_small_large_X_109DW} shows the densities of excess electron and hole for an exciton at the 109\textdegree~domain wall. 
Similar to the case of the 71\textdegree~wall (see Fig.~3 in the main article), \reallynew{a small exciton polaron forms spontaneously}.
\begin{figure}[tb]
  \includegraphics[width=0.5\textwidth]{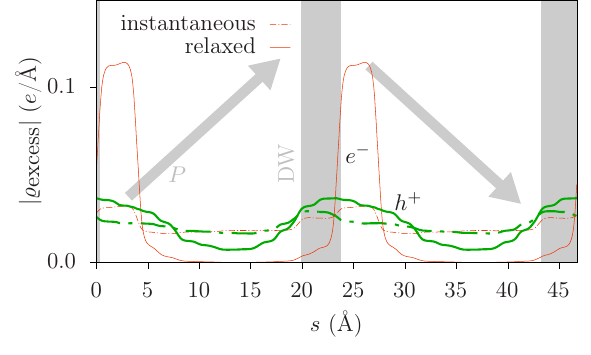}
  \caption{\label{fig:densities_small_large_X_109DW} \new{Smoothened densities of excess electron and hole for 
      excitons ($X$) at the 109\textdegree~domain wall before (``instantaneous'') and after (``relaxed'') polaron formation
	\reallynew{for an $X$ density of 1~$X$ per 120-atom supercell ($\approx3.3\cdot 10^{14}$~$X$/cm$^2$).}} 
  }
\end{figure}
} % end new
%%%%%%%%%%%%%%%%%%%%%%%%%%%%%%
\subsection*{Photovoltage}
%%%%%%%%%%%%%%%%%%%%%%%%%%%%%%
%\begin{figure}
%   \def\svgwidth{0.31\textwidth}
%   \input{DW_structure_109.pdf_tex}
%   \includegraphics[width=0.375\textwidth]{pol_pot_109DW_GS.pdf}
%   \caption{\label{fig:GS}(a) Supercell consisting of 120 atoms with 109\textdegree\ domain walls,
%   (b) three components of the ionic polarization, (c) ionic polarization component $P_s$ across the domain wall, 
% (d) electronic potential energy $V_e^{P_s}$ (thin blue lines) calculated from $P_s$, and the true electronic potential energy $V_e$ (thick black lines), shifted to zero. 
% The potential energy for open-circuit conditions is drawn with dashed lines.}
%\end{figure}

\begin{figure}[htb]
  \includegraphics[width=\columnwidth]{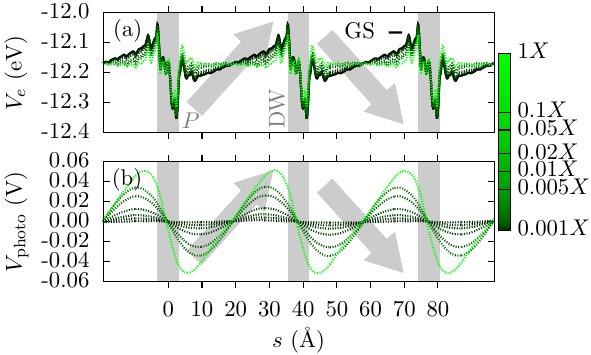}
%  \caption{\label{fig:pot_71_all_280at} (Color online) The same as Fig.~\ref{fig:pot_71_all}, but for a 280-atom supercell.}
  \caption{\label{fig:pot_71_all_280at}(a) Ground-state and excited-state potential profile for exciton densities from 0.001 to 1 excitons ($X$) per \new{280-atom supercell with 71\textdegree\ domain walls}, 
  (b) photovoltage \new{profiles} for \new{these} densities.}
\end{figure}

\begin{figure}[htb]
  \includegraphics[width=\columnwidth]{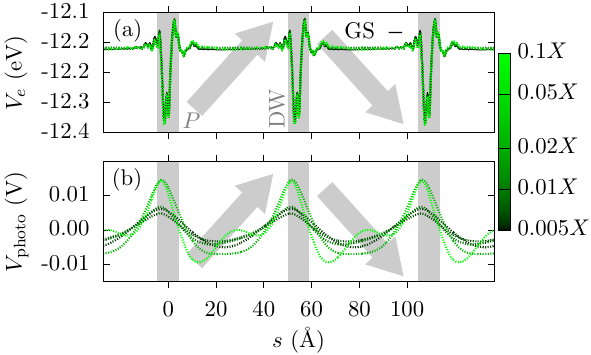}
%  \caption{\label{fig:pot_109_all_280at}The same as Fig.~\ref{fig:pot_109_all}, but for a 280-atom supercell and exciton densities from 0.005~$X$ to 0.1~$X$.}
  \caption{\label{fig:pot_109_all_280at}The same as Fig.~\ref{fig:pot_71_all_280at} for the 109\textdegree~domain wall. Exciton densities range from 0.005~$X$ to 0.1~$X$ per supercell.}
\end{figure}

Figures~\ref{fig:pot_71_all_280at} and 
%\ref{fig:pot_109_all} 
\ref{fig:pot_109_all_280at} 
show all calculated photovoltage profiles for the 71\textdegree\ and the 109\textdegree\ domain wall in the \new{280-atom} supercell.
%\new{In the following, only exciton densities up to 0.1 excitons per supercell are considered, as only these follow a clear trend that can be extrapolated to large domain-wall spacings 
%(see below).}
%
%\begin{figure}[htb]
%  \includegraphics[width=\columnwidth]{pot_71DW_all.pdf}
%  \caption{\label{fig:pot_71_all}(a) Ground-state and excited-state potential profile for exciton densities from 0.001 to 1 excitons ($X$) per \new{120-atom supercell with 71\textdegree\ domain walls}, 
%  (b) photovoltage \new{profiles} for \new{these} densities.}
%\end{figure}
%
%\begin{figure}[htb]
%  \includegraphics[width=\columnwidth]{pot_109DW_all.pdf}
%  \caption{\label{fig:pot_109_all}The same as Fig.~\ref{fig:pot_71_all} for the 109\textdegree\ domain wall.}
%\end{figure}
%

\new{
  The open-circuit (OC) voltage profiles were obtained from the short-circuit (SC) voltage profiles by adding a constant gradient 
  such that in the domain interior the resulting OC voltage slope vanishes, as depicted in Fig.~\ref{fig:VOC_calc_meth}. 
\begin{figure}[htb]
  \includegraphics[width=0.9\columnwidth]{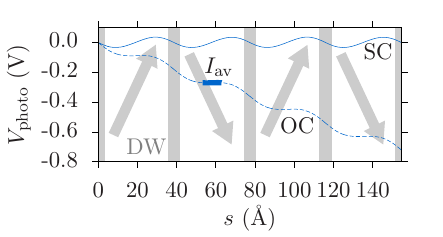}
  \caption{\label{fig:VOC_calc_meth} Short-circuit (SC) photovoltage profile (thin solid line), open-circuit (OC) photovoltage profile (dashed line) 
   for a 280-atom supercell with 71\textdegree\ domain walls,
    and the interval $I_{\mathrm{av}}$ (thick solid line) in which the average voltage slope was compensated by a constant gradient. 
 }
\end{figure}
}

\new{
  Table~\ref{tab:delta_V} contains the calculated short-circuit and open-circuit domain-wall photovoltages obtained with different supercell sizes (different domain-wall distances), 
and those extrapolated to the limit of infinitely large domain-wall distances.
In the case of the 71\textdegree\ domain wall the extrapolated voltages (numbers in \textbf{boldface} in  Tab.~\ref{tab:delta_V}) are depicted in Fig.~7 in the main article. 
In the case of the 109\textdegree\ domain wall it is not possible to perform such an extrapolation because there is no strongly confining 
potential slope. Instead we depict the range of photovoltages between that of the largest supercell (280 atoms), 
and the largest photovoltage as a function of supercell size (numbers in \textbf{boldface} in  Tab.~\ref{tab:delta_V}).
%\todo{SC, OC photovoltages as a function of supercell size, and extrapolated V. Only for the 71 DW. Done.}
%
%
\begin{table}[htb]
      \begin{tabular}{l | l || l | l | l | l || l | l | l | l}
        \multicolumn{2}{c ||}{} & \multicolumn{4}{c||}{71\textdegree~DW} &    \multicolumn{4}{c}{109\textdegree~DW} \\\hline
              &                       & \multicolumn{3}{c|}{$V_{\mathrm{photo}}^{\mathrm{DW}}$ (mV)}  & $d_{\mathrm{DW}}$ & \multicolumn{3}{c|}{$V_{\mathrm{photo}}^{\mathrm{DW}}$ (mV)} &  $d_{\mathrm{DW}}$ \\
        $n_X$ &  $n_{\mathrm{atoms}}$ & SC$_\mathrm{us}$                 & OC$_\mathrm{us}$ & OC      &   (\AA)   & SC$_\mathrm{us}$ & OC$_\mathrm{us}$ & OC &   (\AA)            \\\hline
        0.005 & 120                   & 1.47                             & 3.68          &   0.829    &16.6       &  2.47     &    -0.281    &  -0.0558   & 23.4    \\ 
              & 160                   &                                  &               &            &22.1       &           &              &            & 31.2    \\ 
              & 200                   & 3.86                             & 11.0          &   2.48     &27.6       &  5.77     &     12.4     &   2.46     & 39.0    \\ 
              & 240                   & 5.76                             & 15.1          &   3.40     &33.2       &  6.47     &     15.7     & \bf{3.12}  & 46.8    \\ 
              & 280                   & 7.85                             & 18.1          &   4.08     &38.7       &  11.7     &     5.36     & \bf{1.06}  & 54.6    \\ 
              & $\infty$              &                                  & 23.9          &  \bf{5.38} &$\infty $  &           &              &            &$\infty$ \\ \hline
        0.01  & 120                   & 2.90                             & 7.28          &  1.64      &16.6       & 4.85      &  0.03        &  0.00596   & 23.4     \\ 
              & 160                   & 4.03                             & 10.3          &  2.32      &22.1       &           &              &            & 31.2     \\ 
              & 200                   & 7.16                             & 20.4          &  4.60      &27.6       & 10.9      &  21.0        &  4.17      & 39.0     \\ 
              & 240                   & 10.8                             & 29.1          &  6.56      &33.2       & 14.2      &  26.0        & \bf{5.17}  & 46.8     \\ 
              & 280                   & 14.4                             & 36.0          &  8.11      &38.7       & 8.49      &  6.13        & \bf{1.22}  & 54.6     \\ 
              & $\infty$              &                                  & 52.9          &  \bf{11.9} &   $\infty$&           &              &            & $\infty$  \\ \hline
        0.02  & 120                   & 5.66                             & 14.2          &  3.20      &16.6       & 8.92      &   -0.39      & -0.0775    & 23.4     \\ 
              & 160                   &                                  &               &            &22.1       &           &              &            & 31.2     \\ 
              & 200                   & 13.3                             & 37.5          &  8.45      &27.6       & 17.6      &    35.4      & \bf{7.03}  & 39.0     \\ 
              & 240                   & 20.1                             & 54.7          &  12.3      &33.2       & 20.2      &    33.3      &  6.62      & 46.8     \\ 
              & 280                   & 26.5                             & 68.9          &  15.5      &38.7       & 11.4      &    8.69      & \bf{1.73}  & 54.6         \\ 
              & $\infty$              &                                  & 111           &  \bf{25.0} & $\infty$  &           &              &            &$\infty$  \\ \hline
        0.05  & 120                   & 13.2                             & 33.2          &  7.48      &16.6       & 17.1      &   1.58       &  0.314     & 23.4     \\
              & 160                   &                                  &               &            &22.1       &           &              &            & 31.2     \\
              & 200                   & 28.2                             & 79.1          &  17.8      &27.6       & 28.7      &   50.9       &  10.1      & 39.0     \\
              & 240                   & 41.1                             & 113           &  25.5      &33.2       & 33.6      &   53.8       &  \bf{10.7} & 46.8     \\
              & 280                   & 52.0                             & 138           &  31.1      &38.7       & 22.7      &   3.95       & \bf{0.785} & 54.6         \\
              & $\infty$              &                                  & 186           &  \bf{41.9} & $\infty$  &           &              &            &$\infty$  \\\hline
        0.1   & 120                   & 23.8                             & 59.7          & 13.5       &16.6       &26.3       & -3.04        & 0.604      & 23.4     \\ 
              & 160                   & 28.5                             & 77.0          & 17.3       &22.1       &21.8       &  20.7        & 4.11       & 31.2     \\ 
              & 200                   & 44.2                             & 123           & 27.7       &27.6       &38.0       &  54.9        & \bf{10.9}  & 39.0     \\ 
              & 240                   & 59.4                             & 162           & 36.5       &33.2       &38.2       &  35.4        & 7.03       & 46.8     \\ 
              & 280                   & 69.3                             & 180           & 40.6       &38.7       & 24.2      &  23.4        & \bf{4.65}  & 54.6     \\ 
              & $\infty$              &                                  & 193           & \bf{43.5}  & $\infty$  &           &              &            &$\infty$  \\\hline
        1     & 120                   & 83.8                             &  165          &  37.2      & 16.6      &       &         &        & 23.4     \\ 
              & 160                   & 83.8                             &  173          &  39.0      & 22.1      &       &         &        & 31.2     \\ 
              & 200                   & 90.9                             &  175          &  39.4      & 27.6      &       &         &        & 39.0     \\ 
              & 240                   & 98.1                             &  189          &  42.6      & 33.2      &       &         &        & 46.8     \\ 
              & 280                   & 102                              &  172          &  38.8      & 38.7      &       &         &        & 54.6     
             %& $\infty$              &                                  &               &            & $\infty$  &   &         &        &$\infty$               
      \end{tabular}
      \caption{\label{tab:delta_V} Calculated short-circuit (SC) and open-circuit (OC) domain-wall photovoltages $V_{\mathrm{photo}}^{\mathrm{DW}}$
        for different domain-wall distances $d_{\mathrm{DW}}$ (different numbers of atoms $n_{\mathrm{atoms}}$ in the supercell) 
        and extrapolations to infinitely large domain-wall distances as a function of the number of excitons $n_X$ per supercell for the 71\textdegree\ and the 109\textdegree\ domain walls.
        Unscreened voltages are marked by the subscipt ``us'', such as SC$_\mathrm{us}$.
        The photovoltages printed in \textbf{boldface} are the ones drawn in Fig.~7 in the main article.
      }
     \end{table}
   } % end new

%%%%%%%%%%%%%%%%%%%%%%%%%%%%%%%%%%%%%%%%%%%%%%%%%%%%%%%%%%%
  \subsection{Parameters used in the rate equation}
%%%%%%%%%%%%%%%%%%%%%%%%%%%%%%%%%%%%%%%%%%%%%%%%%%%%%%%%%%%
\new{
  \paragraph{Light penetration depth in a BiFeO$_3$ film}
  The penetration depth calculated from our first-principles absorption coefficient (which is similar to the one measured in Ref.~\onlinecite{zelezny:2010:optical}) is about 33~nm at the photon energy of 3.06~eV that was used in experiment, 
  such that about 95\% of the penetrating light is absorbed in a 100~nm thick film. 
  In Ref.~\onlinecite{alexe:2011:tip} the penetration depth was estimated to be 50~nm, 
  which would result in 86\% of the penetrating light being absorbed.
  \paragraph{Carrier diffusion length}
  It is difficult to accurately estimate the photocarrier diffusion length $l_{\mathrm{diff}}$, therefore two 
  different numbers are considered in the main article. An upper boundary should be the length of a ferroelectric domain.
  \paragraph{Exciton density at the domain wall}
  The planar exciton density $n_X^{\mathrm{DW}}$ is given by the number of excitons $N_X$ and the domain-wall area $A_{\mathrm{DW}}$ 
  contained in the supercell, $n_{X}^{\mathrm{DW}}=N_X/(2A_{\mathrm{DW}})$.
  The domain-wall area is $A_{\mathrm{DW}}\approx42.7$~\new{\AA$^2$\ for the 71\textdegree\ domain wall and $A_{\mathrm{DW}}\approx$~30.2~\AA$^2$ for the 109\textdegree~domain wall.}
  Assuming that all photocarriers within $l_{\mathrm{diff}}$ reach the domain walls, 
  the exciton density can be expressed as $n_X=n_X^{\mathrm{DW}}/l_{\mathrm{diff}}$,
  where $n_X^{\mathrm{DW}}$ is the planar exciton density \new{at the domain wall}.
  Then Eq.~(3) in the main article %\eqref{eq:nX} 
  becomes
  \begin{equation}\label{eq:I_crit}
    I_{\mathrm{light}}=\frac{n_X^{\mathrm{DW}}E_{\mathrm{photon}}d_{\mathrm{film}}}{l_{\mathrm{diff}}(1-R)\tau}.
  \end{equation}
}
\new{
%%%%%%%%%%%%%%%%%%%%%%%%%%%%%%%%%%%%%%%%%%%%%%%%%%%%%%%%%%%
  \subsection{Extrapolation of domain-wall photovoltages to large domain-wall distances}
%%%%%%%%%%%%%%%%%%%%%%%%%%%%%%%%%%%%%%%%%%%%%%%%%%%%%%%%%%%
  Figure~\ref{fig:conv_test_delta_V_71DW} shows the open-circuit photovoltage as a function of the distance between two 71\textdegree~domain walls $d_{\mathrm{DW}}$. 
  The photovoltage does not yet converge for our employed $d_{\mathrm{DW}}$ (supercells with up to 280 atoms), 
  therefore we extrapolate it to the limit of large $d_{\mathrm{DW}}$ by means of a fit function:
  \begin{equation}\label{eq:delta_V_fit}
    V_{\mathrm{photo}}^{\mathrm{DW}}(d_{\mathrm{DW}})=V_{\mathrm{photo}}^{\mathrm{DW}}(\infty)-\Delta V_{\mathrm{photo}}^{\mathrm{DW}}\mathrm{e}^{-b\,d_{\mathrm{DW}}^{3/2}},
  \end{equation}
  \noindent where $V_{\mathrm{photo}}^{\mathrm{DW}}(\infty)$, $\Delta V_{\mathrm{photo}}^{\mathrm{DW}}$, and $b$ are fit parameters. 
  $V_{\mathrm{photo}}^{\mathrm{DW}}(\infty)$ is the extrapolated domain-wall photovoltage that is used in the main article.
  The fit function was chosen based on the exponential decay of the wave function of a particle in a linear potential (Airy function \cite{schleich:2001:quantum}). 
  The data for the largest considered $X$ density of 1\,$X$ per supercell could not be extrapolated to large $d_{\mathrm{DW}}$ and were therefore left out of the analysis.
\begin{figure}[htb]
  \includegraphics[width=0.9\columnwidth]{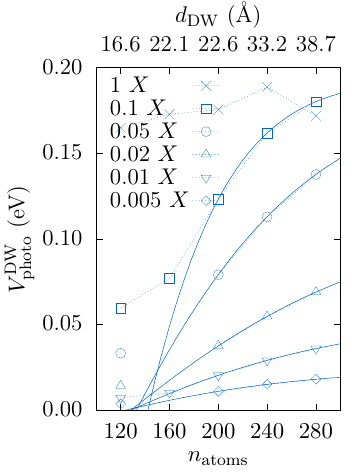}
  \caption{\label{fig:conv_test_delta_V_71DW} Open-circuit photovoltage of the 71\textdegree\ domain wall as a function of the number of atoms $n_{\mathrm{atoms}}$ in the supercell
    and of the distance between domain walls, $d_{\mathrm{DW}}$ (data points from Table~\ref{tab:delta_V}). 
    Solid lines are fits with Eq.~\eqref{eq:delta_V_fit}, dotted lines are a guide to the eye.
  }
\end{figure}
\begin{figure}[htb]
  \includegraphics[width=0.9\columnwidth]{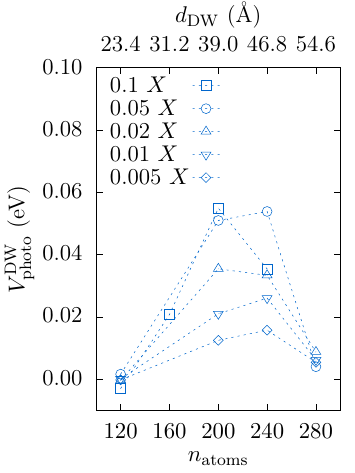}
  \caption{\label{fig:conv_test_delta_V_109DW} The same as Fig.~\ref{fig:conv_test_delta_V_71DW} for the 109\textdegree\ domain wall. Note the change of scale.
  }
\end{figure}
}
\new{
  In the case of the 109\textdegree~domain wall, see Fig.~\ref{fig:conv_test_delta_V_109DW}, the domain-wall photovoltage increases, then decreases with increasing domain-wall distance. 
Here we cannot easily extrapolate to large domain-wall distances, instead we consider the maximum photovoltage as a function of supercell size as an upper limit.
As a lower limit we take the photovoltage of the largest considered domain-wall distance.
}
%\todo{include max PV in Table for 109 degree, like extrapolated PV for 71DW. Or just highlight the numbers actually used by bf text. Done}
\new{
%%%%%%%%%%%%%%%%%%%%%%%%%%%%%%%%%%%%%%%%%%%%%%%%%%%%%%%%%%%
  \subsection{Extrapolation of domain-wall photovoltages to low light intensities}
%%%%%%%%%%%%%%%%%%%%%%%%%%%%%%%%%%%%%%%%%%%%%%%%%%%%%%%%%%%
The photovoltages shown in Fig.~7 in the main article were extrapolated to low exciton densities (low light intensities $I$) with a power law of the form 
\begin{equation}
  V_{\mathrm{photo}}^{\mathrm{DW}}\left(I\right)=V_{\mathrm{photo}}^{\mathrm{DW}}\left(I_0\right) \left(\frac{I}{I_0}\right)^p
\end{equation}
\noindent that was optimized using the three data points corresponding to the lowest light intensities. In the case of the 71\textdegree~domain wall $p\approx 1.14$. 
The photovoltage data of the 109\textdegree\ DW exhibit too much noise to safely extrapolate them to low light intensities.
}
\new{
%%%%%%%%%%%%%%%%%%%%%%%%%%%%%%%%%%%%%%%%%%%%%%%%%%%%%%%%%%%
\subsection{Smoothening of charge densities and potential}
%%%%%%%%%%%%%%%%%%%%%%%%%%%%%%%%%%%%%%%%%%%%%%%%%%%%%%%%%%%
The rapid and strong oscillations of charge densities and electronic potential at atomic nuclei, which obscure variations on a larger scale, 
were smoothened in the following way: 
First the charge density or potential was averaged in a plane parallel to the domain-wall plane [in the $r$-$t$ plane, compare Figs.~1(a) and 2(a) in the main article], 
then a sliding-window average over an $s$ intervall of one atomic plane spacing, like in Ref.~\cite{meyer:2002:ab}, was applied to the excess charge carrier densities.
In order to smoothen the potential a low-pass filter was applied that removed wavelengths up to one atomic plane spacing.
The two smoothening methods should be roughly equivalent, and it is only for historical reasons that one was used for the densities and the other for the potential.
}
\reallynew{
%%%%%%%%%%%%%%%%%%%%%%%%%%%%%%%%%%%%%%%%%%%%%%%%%%%%%%%%%%%
\subsection{Self-trapping at low carrier concentrations}
%%%%%%%%%%%%%%%%%%%%%%%%%%%%%%%%%%%%%%%%%%%%%%%%%%%%%%%%%%%
Charge carriers become spontaneously self-trapped for all
carrier concentrations considered. Figure 10 shows the iso-
surface of the charge density of an excess electron in a 320-
atom supercell, the lowest carrier concentration considered.
\begin{figure}[htb]
  \includegraphics[width=0.9\columnwidth]{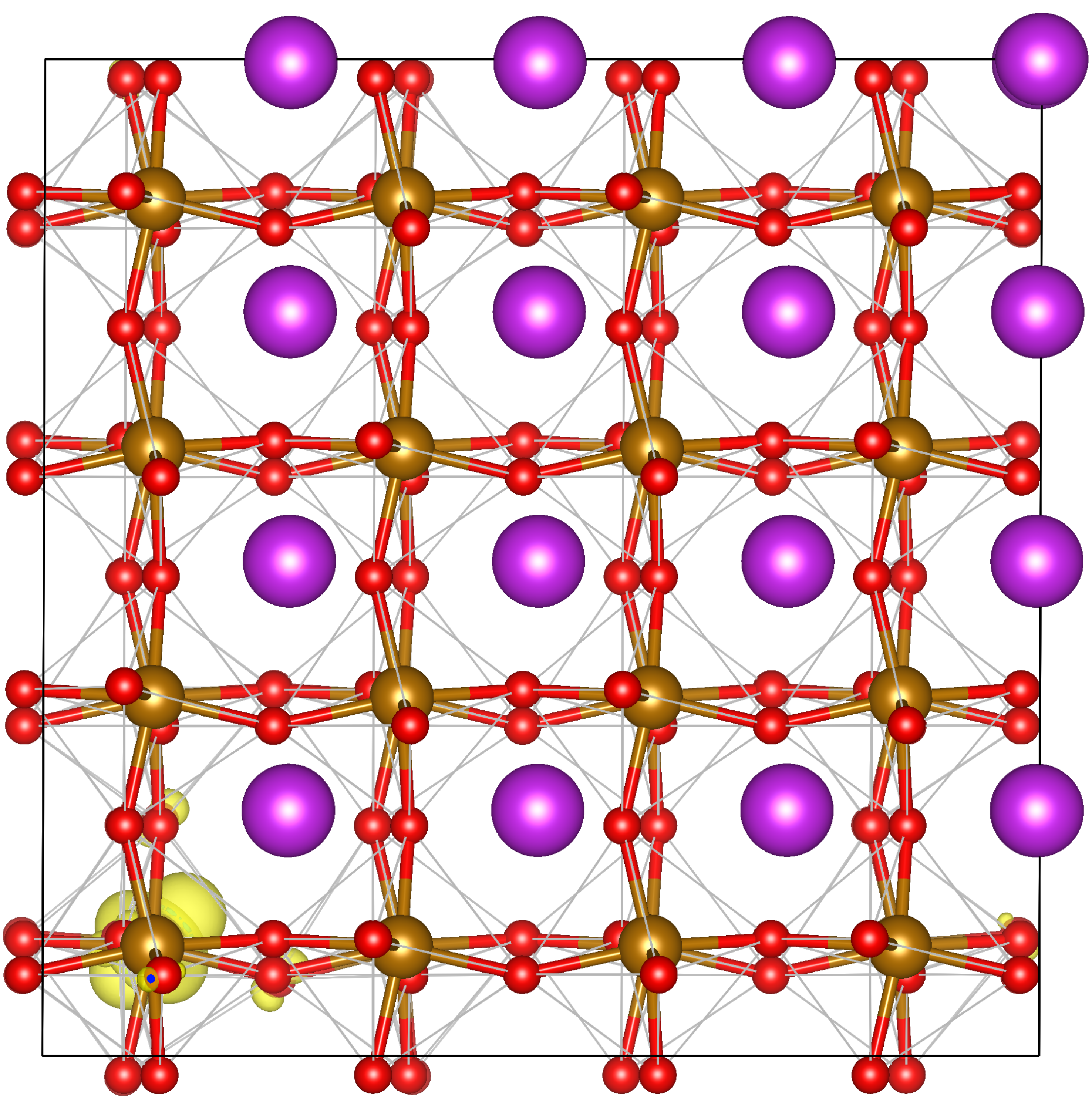}
  \caption{\label{fig:SEP_320_atoms} Self-trapped electron in a 320-atom supercell (yellow:
charge density iso-surface of the exceess electron, large purple
spheres: Bi, medium-sized brown spheres: Fe, small red spheres:
O).
  }
\end{figure}
}
%Figure~\ref{fig:pot_71_relaxed} shows the calculated photovoltage profile including ionic screening for the 71\textdegree\ domain wall.
%In the case of ionic screening, the induced photovoltage consists of a large component that is symmetric with respect to the domain wall plane and 
%a small component that is antisymmetric with respect to the domain wall plane. Only the antisymmetric component is relevant for the macroscopic 
%photovoltage. Because the ionic positions now differ between the ground state and the excited state, 
%the photovoltage profile exhibits a large noise, which we smoothened with a bandpass filter.
%\begin{figure}[htb]
%  \includegraphics[width=\columnwidth]{pot_71DW_0_1X_relaxed.pdf}
%  \caption{\label{fig:pot_71_relaxed}(a) Ground-state and excited-state potential profile for an exciton densities of 0.1 $X$ per supercell, 
%  (b) photovoltage $V_{\mathrm{photo}}$, smoothened photovoltage $V_{\mathrm{photo}}^{\mathrm{sm}}$  and 
%symmetric ($V_{\mathrm{photo}}^{\mathrm{sm, s}}$) and antisymmetric ($V_{\mathrm{photo}}^{\mathrm{sm, as}}$) components of the photovoltage for the 71\textdegree\ domain wall.
%The antisymmetric component is used to calculate the domain-wall photovoltage $V_{\mathrm{photo}}^{\mathrm{DW}}$.}
%\end{figure}

%\clearpage
\bibliography{jabbr,all}